\documentclass[preprint,12pt]{elsarticle}
\usepackage[margin=1in,footskip=0.25in]{geometry}
\usepackage{amsthm}%
\usepackage{mathtools}
\usepackage{amsfonts}%
\usepackage{pdfpages}
\usepackage{color}
\newtheorem{theorem}{Theorem}
\newtheorem{definition}[theorem]{Definition}
\begin{document}

\title{Unusual dynamics and hidden attractors of the Rabinovich-Fabrikant system}

\author[rm1,rm2]{Marius-F. Danca}
\author[rm3,rm4]{Nikolay Kuznetsov\corref{cor1}}
\cortext[cor1]{Corresponding author}
\author[rm5]{Guanrong Chen}
\address[rm1]{Dept. of Mathematics and Computer Science, Avram Iancu University of Cluj-Napoca, Romania}
\address[rm2]{Romanian Institute of Science and Technology, Cluj-Napoca, Romania}
\address[rm3]{Department of Applied Cybernetics, Saint-Petersburg State University, Russia}
\address[rm4] {Department of Mathematical Information Technology, University of Jyv\"{a}skyl\"{a}, Finland}
\address[rm5]{Department of Electronic Engineering, City University of Hong Kong, Hong Kong, China}

\begin{abstract}This paper presents some unusual dynamics of the Rabinovich-Fabrikant system, such as ``virtual'' saddles, ``tornado''-like stable cycles and hidden chaotic attractors. Due to the strong nonlinearity and high complexity, the results are obtained numerically with some insightful descriptions and discussions.
\end{abstract}
\maketitle
\vspace{3mm}

\section{Introduction}

The Rabinovich-Fabrikant (RF) system, introduced in 1979 by Rabinovich and Fabrikant \cite{raba} and numerically investigated in \cite{dancay}, was initially designed as a physical model describing
the stochasticity arising from the modulation instability in a non-equilibrium dissipative medium. However, recently in \cite{dancay} we revealed that beside the physical properties described in \cite{raba}, the system presents some extremely rich dynamics. More significantly, unlike other nonlinear chaotic systems containing only second-order nonlinearities (such as the Lorenz system), the RF system with third-order nonlinearities, presents some unusual dynamics such as ``virtual'' saddles beside several chaotic attractors with different shapes and hidden chaotic attractors. In fact, the interest in this system is continuously increasing (see e.g. \cite{yong,zhan,mots,agra1,chav,umo,ser,sri}).

The mathematical RF model is defined by the following system of ODEs:

\begin{equation}
\label{rf}
\begin{array}{l}
\overset{.}{x}_{1}=x_{2}\left( x_{3}-1+x_{1}^{2}\right) +ax_{1}, \\
\overset{.}{x}_{2}=x_{1}\left( 3x_{3}+1-x_{1}^{2}\right) +ax_{2}, \\
\overset{.}{x}_{3}=-2x_{3}\left( b+x_{1}x_{2}\right),
\end{array}%
\end{equation}

\noindent where $a,b$ are two real positive parameters. The system dynamics depend sensibly on the parameter $b$ but less on $a$. Therefore, $b$ will be considered as the bifurcation parameter\footnote{Negative values also generate interesting dynamics, but this case has not much physical meaning therefore is not considered in the present paper.}.

Due to the extreme complexity of the system (\ref{rf}), a detailed analytical study concerning for example the existence of invariant sets, and of the heteroclinic or homoclinic orbits and so on, is literally impossible. Therefore, this paper takes a numerical analysis approach, to carefully investigate some new complex dynamics, and find hidden attractors of the system, deepening the investigations in \cite{dancay}.

The system presents the following symmetry:
\begin{equation}\label{sim}
T(x_1,x_2,x_3)\rightarrow(-x_1,-x_2,x_3),
\end{equation}

\noindent which means that each orbit has its symmetrical (``twin'') orbit with respect to the $x_3$-axis under transformation $T$, with five equilibria: $X_0^*(0,0,0)$ and
\begin{equation}
\label{eq}
\begin{array}{l}
X_{1,2}^{\ast }\left( \mp \sqrt{\dfrac{bR_1+2b}{4b-3a}},\pm \sqrt{b\dfrac{4b-3a}{R_1+2}},\dfrac{aR_1+R_2}{\left(4b-3a\right) R_1+8b-6a}\right),
\\
X_{3,4}^{\ast }\left( \mp \sqrt{\dfrac{bR_1-2b}{3a-4b}},\pm \sqrt{b\dfrac{4b-3a}{2-R_1}},\dfrac{aR_1-R_2}{\left(
4b-3a\right) R_1-8b+6a}\right) ,
\end{array}%
\end{equation}

\noindent where $R_1=\sqrt{3a^{2}-4ab+4}$ and $R_2=4ab^{2}-7a^{2}b+3a^{3}+2a$.

Extensive numerical tests lead to a conclusion that the available numerical methods for ODEs, such as those implemented in different software packages, might give unexpectedly different results of the system even for the same parameter values and initial conditions. On the other hand, fixed-step-size schemes (such as the standard Runge-Kutta method RK4, or the predictor-corrector LIL method \cite{dancaz} utilized in this paper) can give more accurate results. However, the numerical results depend drastically on the step-size and the initial conditions.

In this paper, it is numerically demonstrated that the RF system presents several unusual ``virtual'' saddle-like which apparently exist for relatively large intervals of $b$ compared with our previous results in \cite{dancay}. Moreover, the existence of hidden attractors is numerically revealed and investigated.

The paper is organized as follows: Section 2 unveils unusual dynamics of the RF system, Section 3 presents two hidden attractors of the RF system with conclusion ending the paper.

\section{Unusual dynamics of the RF system}\label{din}
\begin{figure}
\begin{center}
\includegraphics[clip,width=0.6\textwidth] {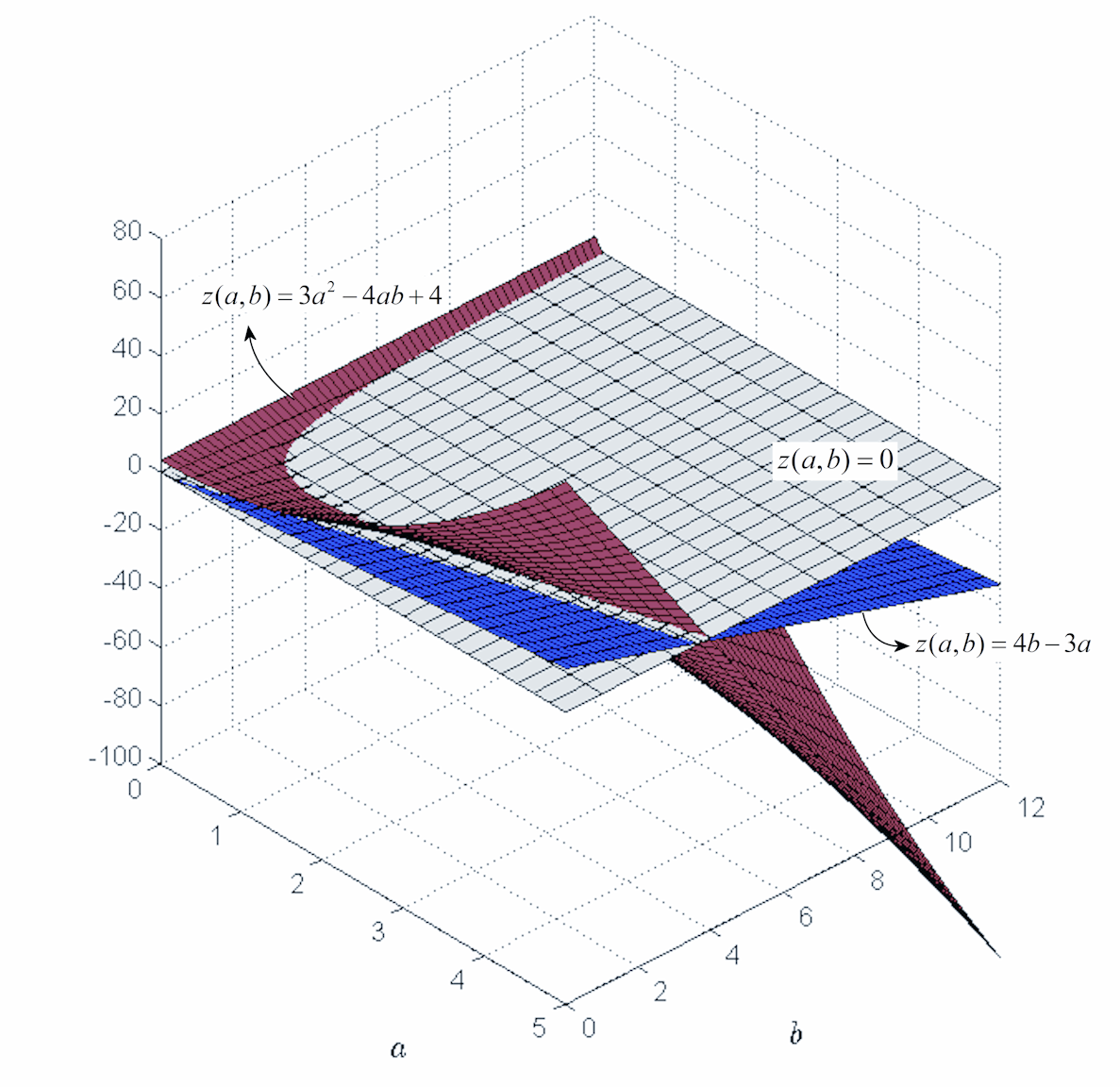}
\caption{Surfaces $z(a,b)=4b-3a$ and $z(a,b)=3a^2-4ab+4$.}
\label{fig0}
\end{center}
\end{figure}

\begin{figure}
\begin{center}
  \includegraphics[clip,width=0.6\textwidth] {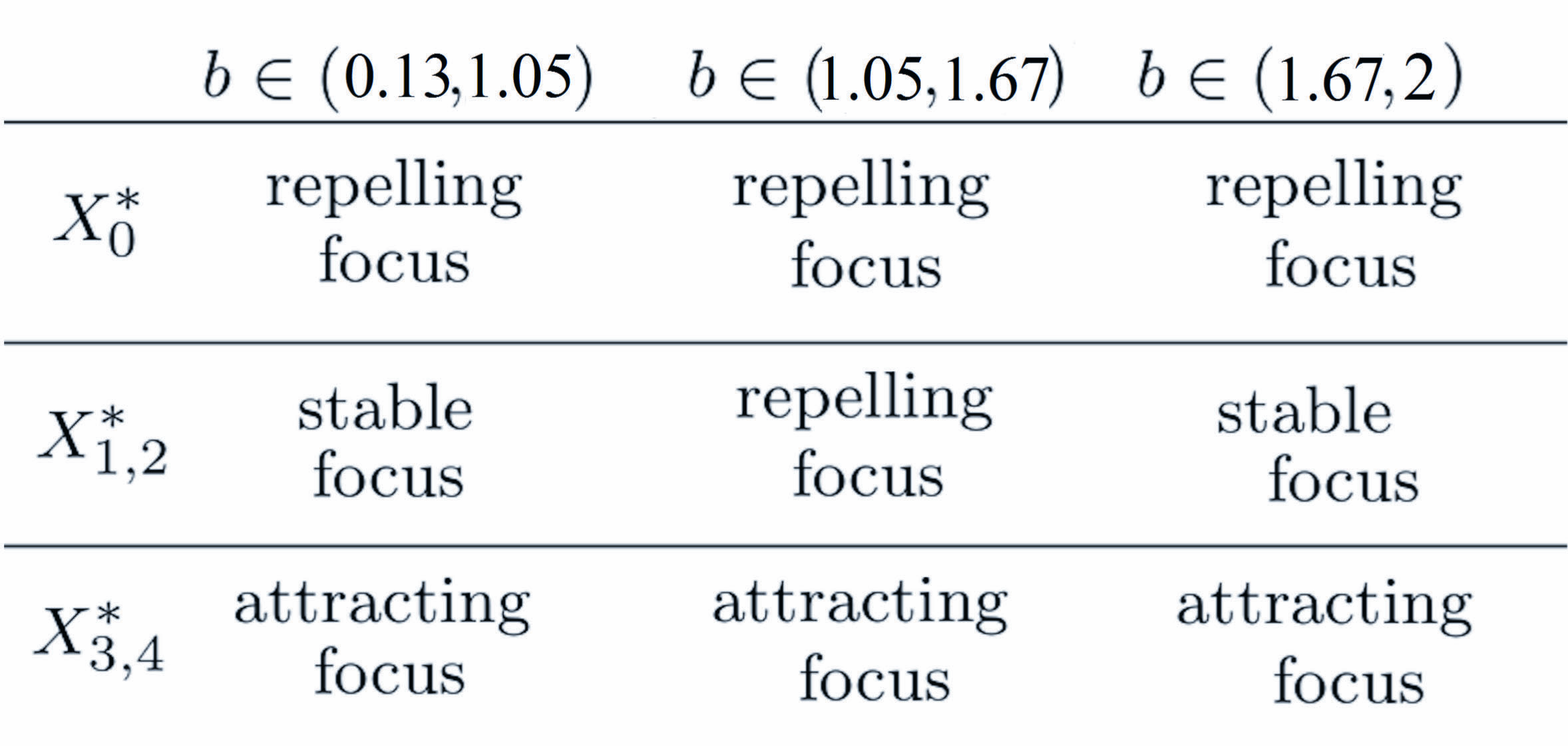}
\caption{Equilibria $X_{0}^*$ and $X_{1,2,3,4}^*$.}
\label{tabel}
\end{center}
\end{figure}

The equilibrium $X_0^*$ exists for all values of $a,b$. The existence of the other equilibria $X_{1,2,3,4}^*$ is determined by the signs of the expressions $4b-3a$ and $3a^2-4ab+4$ (Fig. \ref{fig0}). The stability type is presented in Fig. \ref{tabel}.

\begin{figure}
\begin{center}
\includegraphics[clip,width=0.6\textwidth]{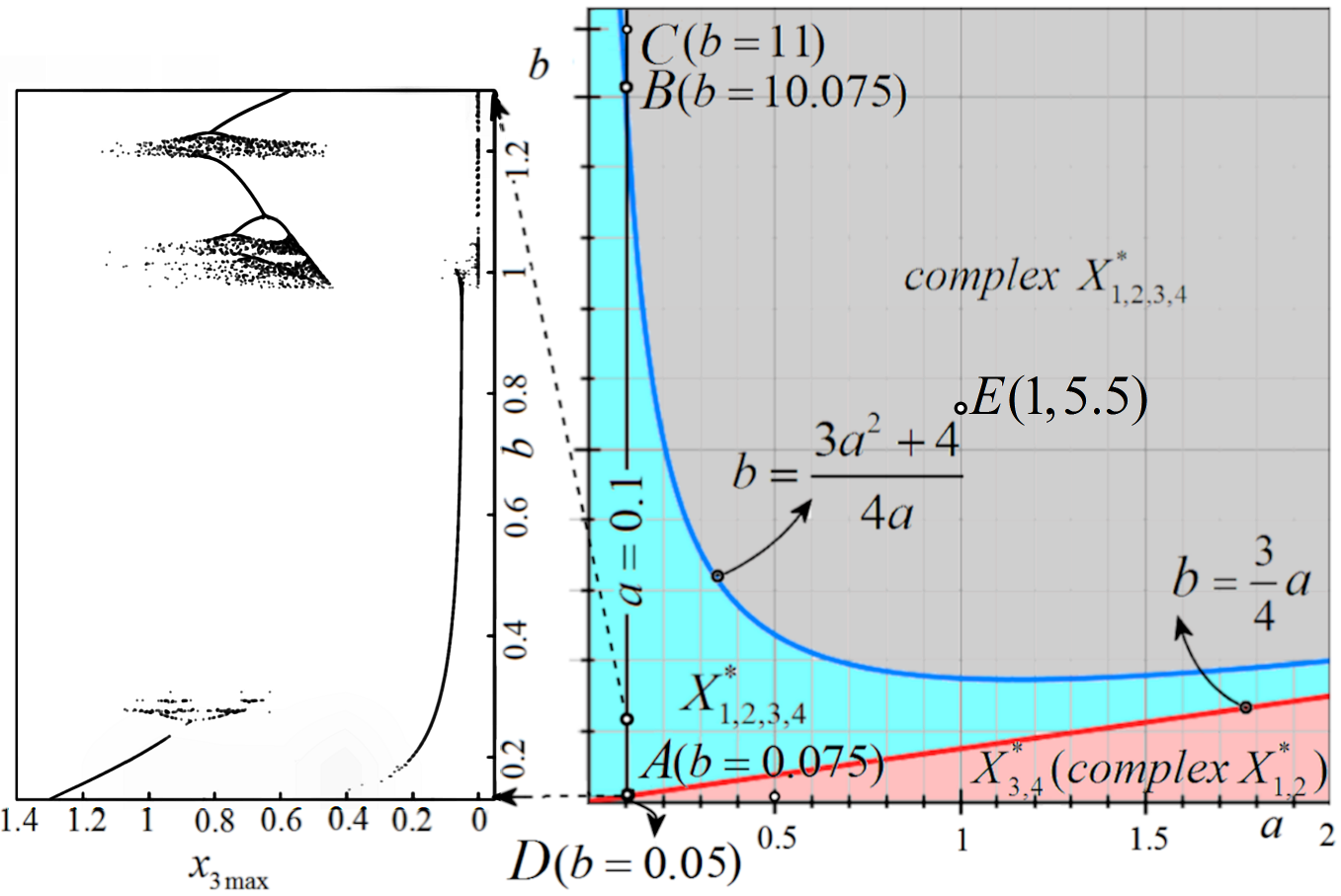}
\caption{Existence domain of equilibria in the parameter plane $(a,b)$ and the bifurcation diagram for $b\in(0.075,1.3)$.}
\label{fig1}
\end{center}
\end{figure}

For equilibria $X_{1,2,3,4}^*$ there exist regions where $X_{1,2,3,4}^*$ does not exist, regions where there exist only equilibria $X_{3,4}^*$ and regions with all equilibria $X_{1,2,3,4}^*$ (Fig. \ref{fig1}). This fact suggests an important reason for having some unusual dynamical behavior in this system.

This paper reports, for the first time, the existence of some unusual dynamics of the RF system, namely the presence of ``virtual'' saddles and the dynamics of ``double-vortex tornado''-like attractors as well as hidden chaotic attractors.

The parameter $b$ is considered as the bifurcation parameter (see Fig. \ref{fig1}, where the bifurcation diagram of $x_3$ is plotted for $b\in(0.05,1.3)$). The parameter $a$ is chosen as $a=0.1$ because for this value one obtains a large domain of existence of all equilibria (segment $AB$ with $A(0.1,b_1)$ and $B(0.1,b_2)$ where $b_1=0.075$, $b_2=10.075$, determined by the intersection of $a=0.1$ and the curves $b=3/4a$ and $b=(3a^2+4)/(4a)$, respectively). For $a>0.1$ and $b>11$, the system becomes generally unstable.

The numerical integrations and underlying computer simulations are obtained with the LIL algorithm \cite{dancaz}, with step-size $h=0.0001-0.005$, while integration time interval is $I=[0,T_{max}]$ with $T_{max}=300-500$. The initial conditions $S=(x_{0,1},x_{0,2},x_{0,3})$, which are of a major impact on the numerical results (especially in finding hidden attractors), are chosen as  $x_{0,1}, x_{0,2}\in (-1,1)$, and $x_{03}\in(0-0.2)$\footnote{For $x_3=0$, the system is unstable \cite{dancay}.}.

\begin{figure}
\begin{center}
\includegraphics[clip,width=0.6\textwidth]{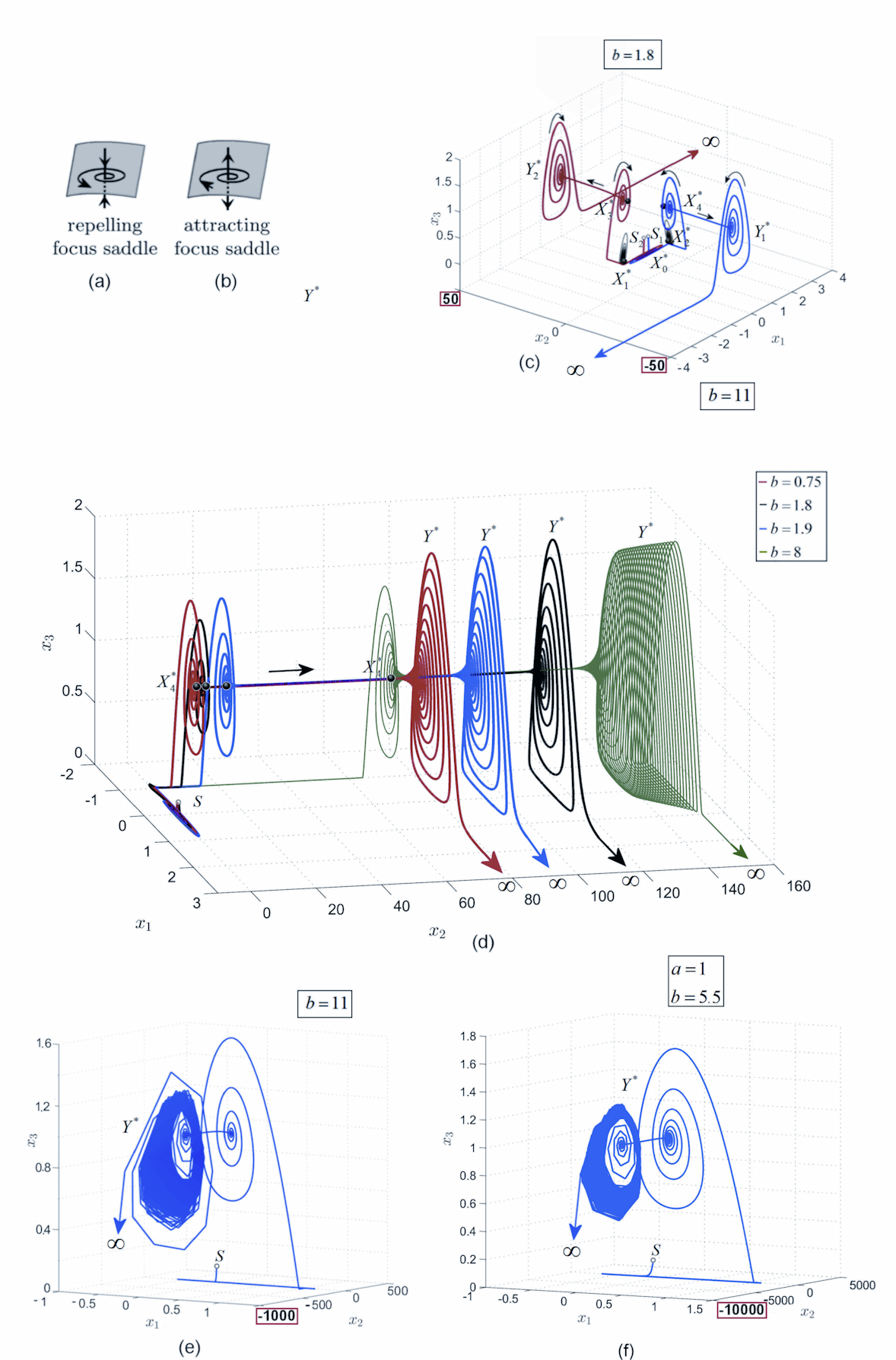}
\caption{(a), (b) Repelling and attracting focuses (sketch). (c) ``Virtual'' saddle for $b=1.8$ \cite{dancay}; $S_{1,2}$ are initial conditions. (d) ``Virtual'' saddles for $b=0.75, 1.8, 1.9, 8$. (e) ``Virtual'' saddle for $b=11$ (equilibria $X_{3,4}^*$ does not exist). (f) ``Virtual'' saddle for $a=1$ and $b=5.5$ (equilibria $X_{3,4}^*$ does not exist).}
\label{fig3}
\end{center}
\end{figure}

For $b=1.8$, in \cite{dancay} a singular but interesting behavior was found, which suggests a kind of ``virtual'' reppeling focus saddles, denoted by $Y^*$ (Fig. \ref{fig3} (c)-(e)). As is well known, usually a fixed point is a reppeling or attracting focus, or a saddle. In this case, the dynamics in its neighborhood look like the sketch in Figs. \ref{fig3} (a), (b), respectively (see also Fig. \ref{fig3} (c), where the pair of saddles $X_{3,4}^*$ are plotted). However, as can be seen in Figs. \ref{fig3} (d)-(e), beside the singular case found in \cite{dancay} (Fig. \ref{fig3} (b)), it can be seen that this behavior is a generic behavior for all values of $b\in(b_1,b_2)$ (see Fig. \ref{fig3} (d), where for the sake of image clarity only the saddle $X_4^*$ is plotted). It is interesting to observe that for $b$ values lying on the segment $AB$ (situated in the existence domain of all equilibria, Fig. \ref{fig1}), these saddle-like points are connected by smooth curves to the real saddles $X_{3,4}$ appearing as a reversed ``replica'' of $X_{3,4}$ and being emanated by them.

These saddle-like points exist not only in the existence domain of all equilibria (segment $AB$), but also in some nonexistence domain such as the point $C (b=11)$ (Fig. \ref{fig1}). While, for $b\in AB$, the ``virtual''-saddles seem to appear as a consequence of the existence of the real saddles $X_{3,4}^*$, in the nonexistence domain of the equilibria. Wherein the ``virtual''-saddles still appear, even $X_{3,4}^*$ does not exist, which are complex along the $x_2$ axis (Fig. \ref{fig1}).

Another interesting characteristic of these saddles-like points $Y^*$ is that in the existence domain of the real saddles $X_{3,4}^*$, depending on $b$, the distance between them and $X_{3,4}^*$ exceeds several times of the underlying attractor size (Figs. \ref{fig3} (c)-(e)). The unstable direction of $X_{3,4}^*$ acts as stable direction for $Y^*$ and is the same for each $b$. Also, the spinning orientation on $Y_{1,2}^*$ is the same as that on $X_{3,4}$ (Figs. \ref{fig3} (c) and (d)).

Compared to the case $b=0.2715$, where the saddles-like point $Y_1^*$ (Figs. \ref{ffig4} (a) and (b)) is obtained by starting the numerical integration with initial points in the neighborhoods of $X_4^*$ (similarly for $Y_2^*$), in the case of $b=0.2876$, $Y_1^*$ (Figs. \ref{ffig3} (a) and (b)) is generated either by starting from the neighborhood of $X_0^*$ (gray trajectory) or by starting from the neighborhood of $X_4^*$ (dark-brown trajectory) (similarly for $Y_2^*$). Because $Y_{1,2}^*$ are generated from points in the vicinity of unstable equilibria $X_0^*$ and $X_{3,4}$, and because the divergence to infinity indicated by the numerical tests, these points can be classified as unbounded self-excited attractors.

\begin{figure}
\begin{center}
\includegraphics[clip,width=0.7\textwidth]{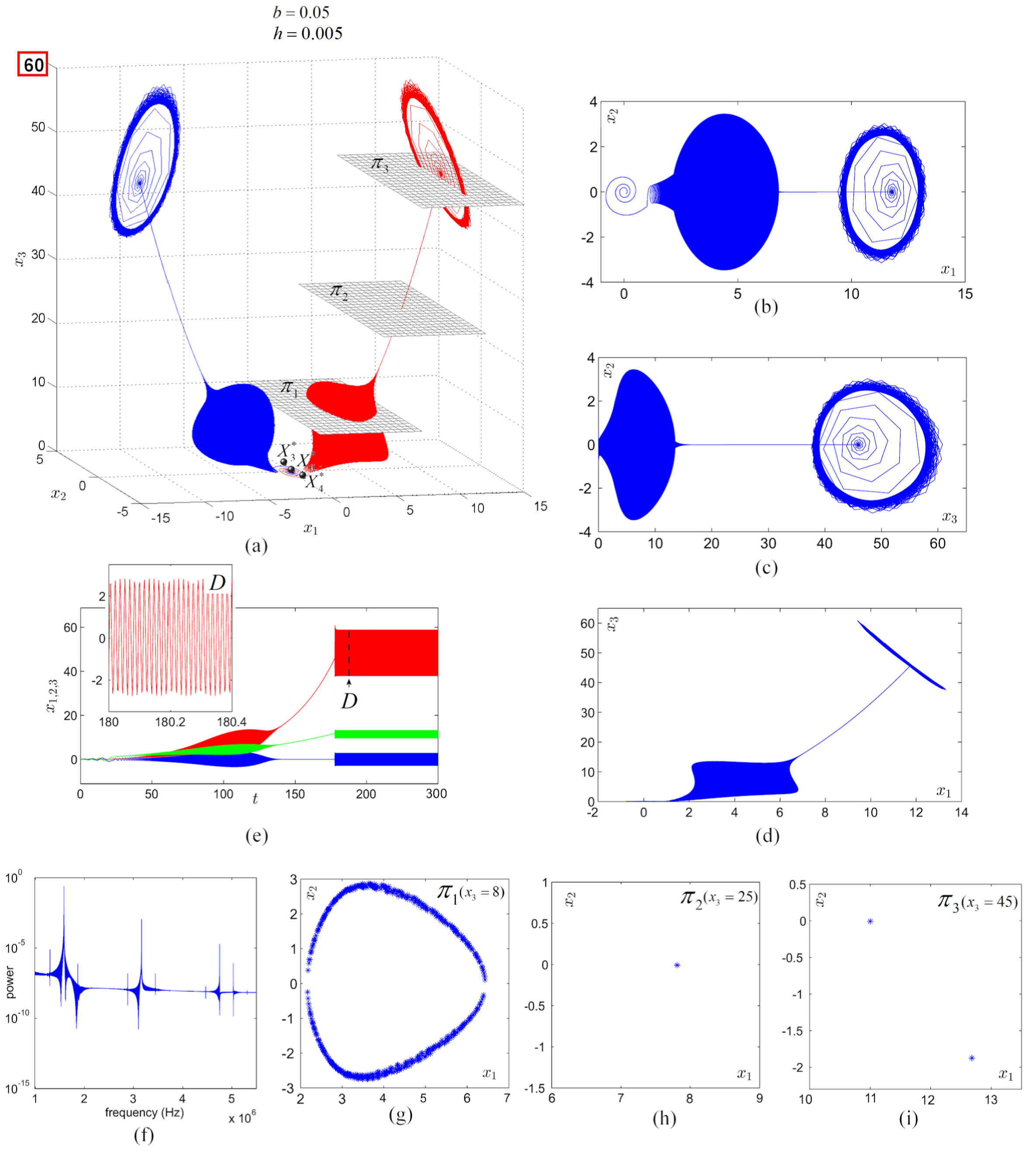}
\caption{(a) Double-vortex-like ``tornados'' for $b=0.05$ and $h=0.005$. (b)-(d) Planar phase plots. (e) Time series and inset $D$. (f) Power spectrum. (g)-(i) Poincar\'{e} sections.}
\label{fig4}
\end{center}
\end{figure}

For values of $b<0.075$ (point $D (b=0.05$), Fig. \ref{fig1}) and $h=0.005$, these saddle-like points disappear and transform into stable ``double-vortex tornado''-like cycles (Fig. \ref{fig4} (a)). Plane phase plots (Figs. \ref{fig4} (b)-(d)), time series, the inset $D$ (Fig. \ref{fig4} (e)), the power spectrum (Figs. \ref{fig4} (f)) and Poincar\'{e} sections (Figs. \ref{fig4} (g)-(i)), all together show a possible cvasiperiodic motion. It is noted that these attractors exist only in the regions where equilibria $X_{3,4}^*$ exist.

\begin{figure}
\begin{center}
\includegraphics[clip,width=0.35\textwidth]{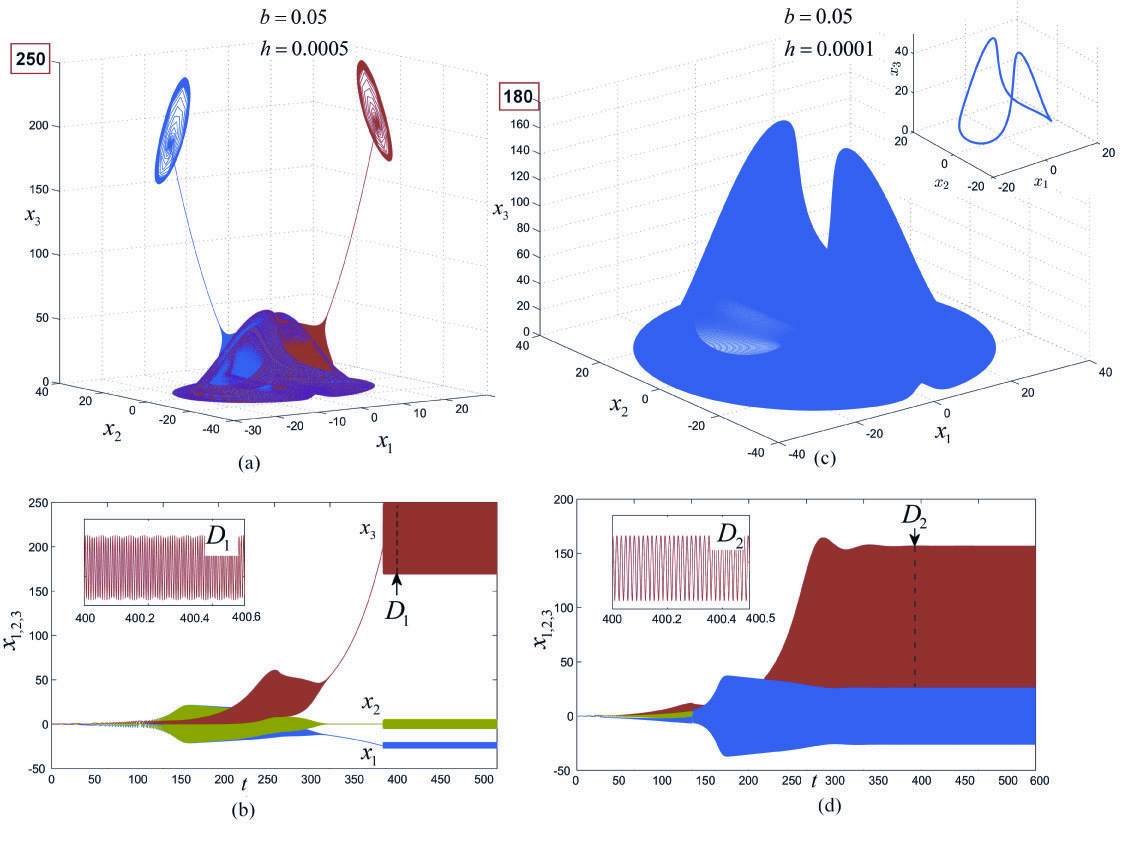}
\caption{ (a) Double-vortex-like ``tornados'' for $b=0.05$ and $h=0.0005$. (b) Time series and inset $D_1$. (c) Stable cycles with and without transients for $b=0.05$ and $h=0.0001$, respectively. (d) Time series and inset $D_2$.}
\label{fig5}
\end{center}
\end{figure}

As mentioned before, spectacular phenomena appear depending on the $h$ size. In fact, for a smaller integration step-size, $h=0.0005$, the vortexes still exist, but with a larger size along the $x_3$ axis along with an enlarged base (see the phase plot in Fig. \ref{fig5} (a) and the time series with inset $D_1$ in Fig. \ref{fig5} (b)), while for $h=0.0001$, the vortices disappear but their base seems to transform into a stable periodic motion (see the phase plot with and without transients in Fig. \ref{fig5} (c), and the time series and the inset $D_2$ in Fig. \ref{fig5} (d)).


\section{Hidden chaotic attractors of the RF system}
From a computational perspective, it is natural to suggest the following classification of attractors\footnote{ Since from a computational perspective, it is essential that an attractor can be found numerically, we adopt the following definition:
A closed bounded invariant set $K$ is called an attractor if there exists an open neighborhood of $K$: $K_\varepsilon \supset K$, such that for all initial data form $K_\varepsilon$ (or except a zero measure set) the corresponding trajectories tend to $K$  as $t$ tends to $+\infty$.
This gives hope that the attractor can be revealed numerically, despite the possible computational error (we do not consider the computational difficulties caused by the shape of basin of attraction, e.g. by Wada and riddled basins).
E.g., a semistable trajectory on the plane has the basin of attraction with positive measure, but can hardly be computed by the usual methods because of the discretization step in the computational procedures (see, e.g. \cite{ixus1,ixus2,ixus3}).},
which is based on the connection of their basins of attraction with equilibria in the phase space

\begin{definition}\label{def}\cite{unu,doi,trei,patru}
An attractor is called a self-excited attractor if its basin of attraction intersects with any open neighborhood of a stationary state (an equilibrium); otherwise, it is called a hidden attractor.
\end{definition}

Self-excited attractors can be visualized numerically by a standard computational procedure, in which after a transient process a trajectory, starting from a point of a neighborhood of unstable equilibrium, attracted to the attractor.
In contrast, the basin of attraction for a hidden attractor is not connected with any equilibrium and, thus, for the numerical localization of hidden attractors it is necessary to develop special analytical-numerical procedures in which an initial point can be chosen from the basin of attraction. For example, hidden attractors are attractors in systems with no-equilibria or in multistable systems with only stable equilibrium.
At the same time the coexisting self-excited attractors in multistable systems (see, e.g. various examples of multistable engineering systems in \cite{sapte}, and recent physical examples in \cite{opt}) can be found using a standard computational procedure, whereas there is no regular way to predict the existence or coexistence of hidden attractors in a system.

\begin{figure}
\begin{center}
  \includegraphics[clip,width=0.85\textwidth] {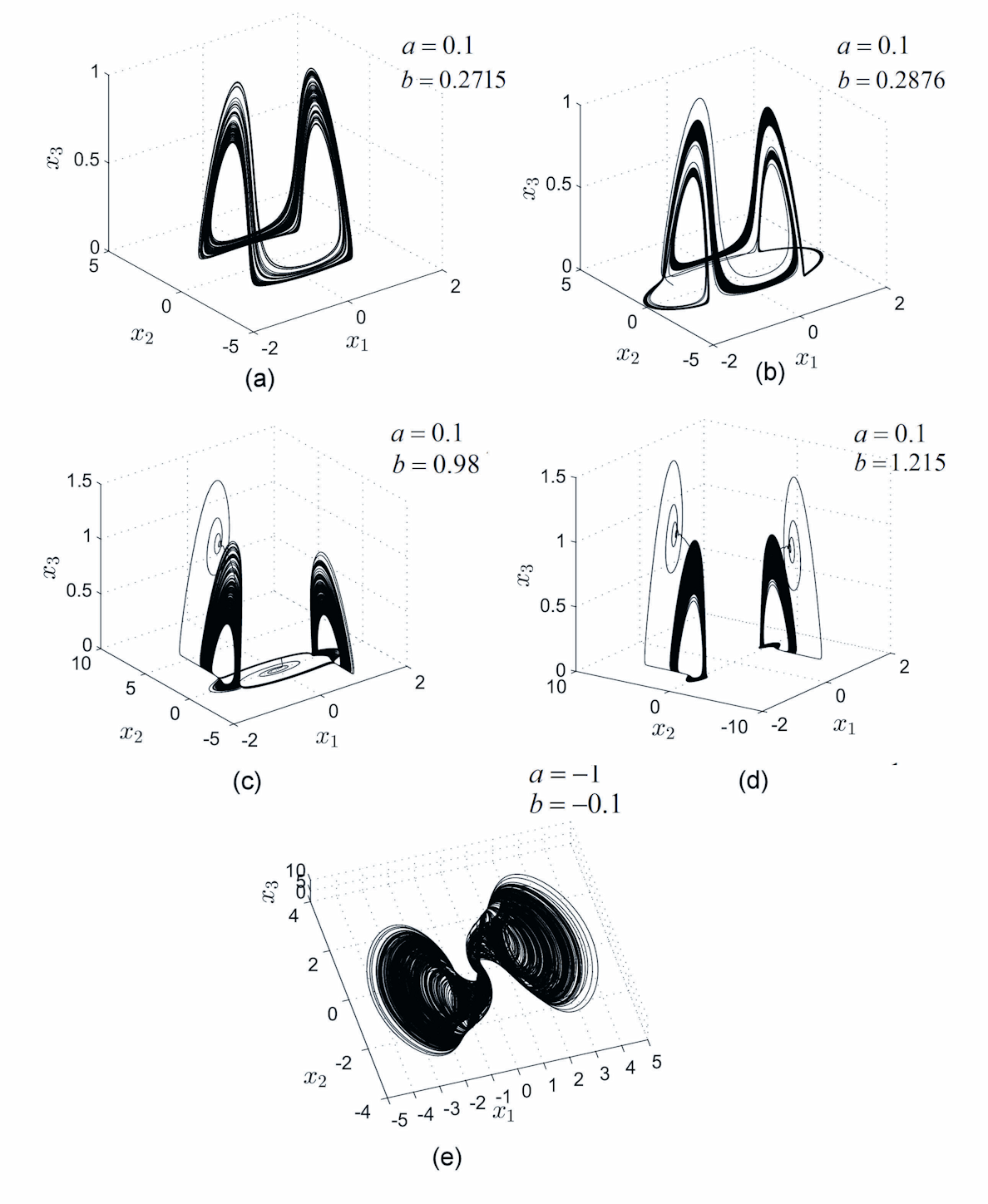}
\caption{Chaotic attractors. (a), (b) Hidden chaotic attractors. (c)-(e) Self-excited chaotic attractors.}
\label{ffig2}
\end{center}
\end{figure}

Among the complicated and unusual dynamics analyzed in Section \ref{din} (such as multistability (coexistence of chaotic attractors and stable cycles), heteroclinic orbits connecting different kinds of attractors \cite{dancay}, the system also presents several chaotic attractors with different shapes (Fig. \ref{ffig2}). Among these attractors, it is found that two attractors, corresponding to $b=0.2876$ and $b=0.2715$ respectively, are hidden. The others are self-excited attractors: there does not seem to exist small neighborhoods around the unstable equilibria such that all trajectories starting inside these neighborhoods tend to infinity or are attracted by stable equilibria.

Now, consider the chaotic attractor $H_1$ (Fig. \ref{ffig2} (b)) corresponding to $a=0.1 \quad \text{and}\quad b=0.2876$. In order to show that this attractor is hidden, Definition \ref{def} there does not seem to exist that its attraction basin does not intersect with small neighborhoods of unstable equilibria. For this purpose, initial points are chosen inside some $\delta$-vicinity of unstable equilibria (here $\delta= 0.05$) and the system is integrated to see if the obtained trajectories are attracted by the chaotic attractor. For all unstable equilibria, this procedure is repeated for several times (with $100$ random choice here, to get different initial points).

For the chosen parameters, the equilibrium $X_0^*$ is a repelling focus-saddle, equilibria $X_{1,2}^*$ are stable focus-nodes and equilibria $X_{3,4}^*$ are attracting focus saddle (see \cite{dancay} and Table \ref{tabel}).

\begin{figure}
\begin{center}
  \includegraphics[clip,width=0.85\textwidth] {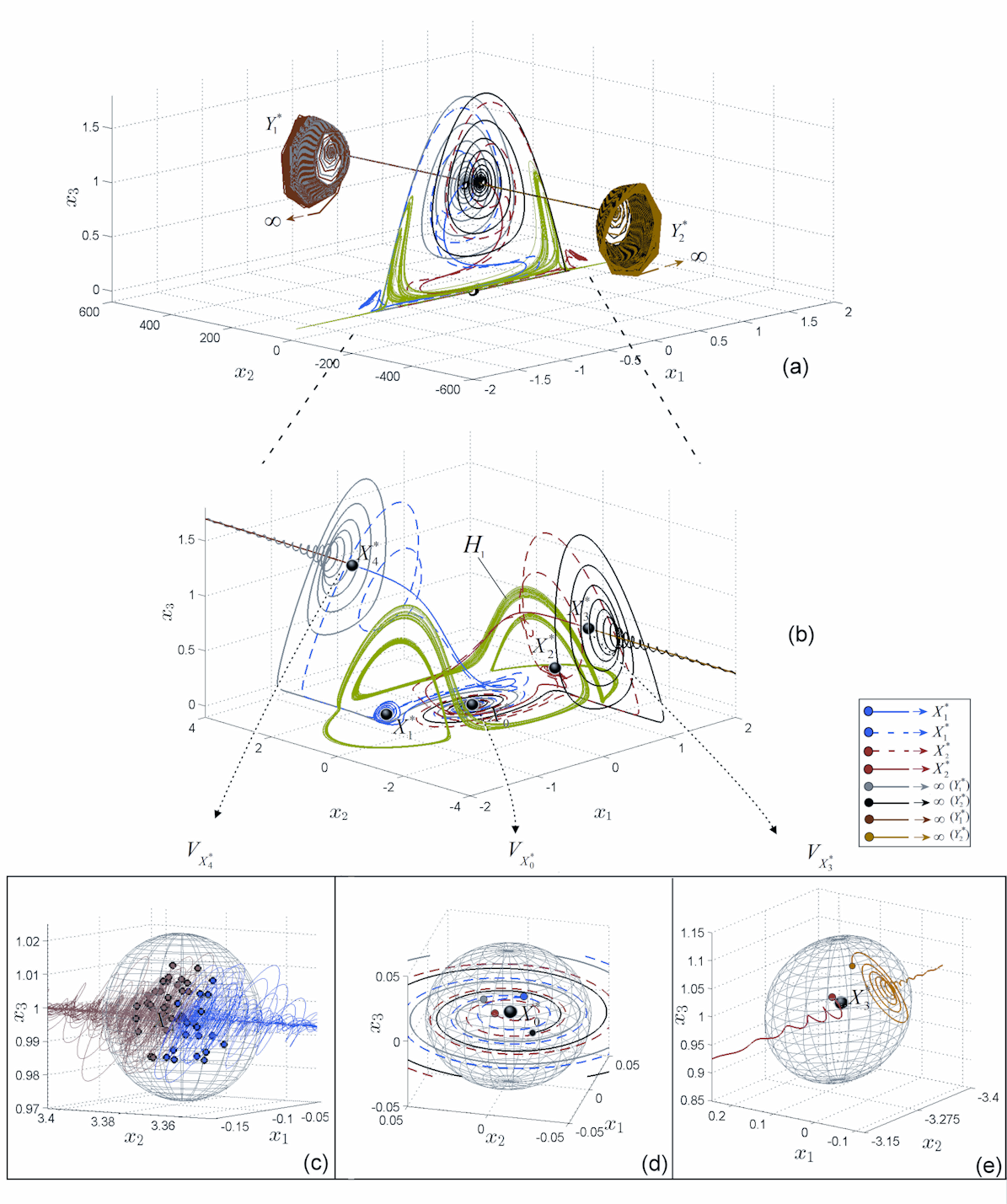}
\caption{(a) Hidden chaotic attractor (green) for $b=0.2876$ and related ``virtual'' saddles $Y_{1,2}^*$. (a) Detail of the hidden chaotic attractors, trajectories diverging to infinity (via $Y_{1,2}^*$) and trajectories attracted by the stable equilibria $X_{1,2}^*$. (c) Trajectories starting from the $\delta$-vicinity of $X_4^*$ either are attracted by the equilibrium $X_1^*$ (blue), or diverge to infinity via $Y_1^*$ (dark brawn) (50 trajectories). (d) Trajectories starting from the $\delta$-vicinity of $X_0^*$, either diverge to infinity via $Y_{1,2}^*$ (black and grey trajectories), or tend to $X_{1,2}^*$ (dotted red and blue trajectories) (4 representative trajectories). (e) Trajectories starting form the $\delta$-vicinity of $X_3^*$ either tend to $X_2^*$ (red), or diverge to infinity (brawn) via $Y_2^*$ (two representative trajectories).}
\label{ffig3}
\end{center}
\end{figure}

First, check the $\delta$-vicinity of the equilibrium $X_0^*$, $V_{X_0^*}$. As proved analytically in \cite{dancay}, for $x_3=0$, all trajectories starting from $V_{X_0^*}$ diverge to infinity and therefore are not considered in these numerical simulations. The origin is globally asymptotically unstable. For $x_3>0$ (as required by the system physical structure \cite{raba}), the numerical simulations show that all trajectories starting from $V_{X_0^*}$ either tend to $\infty$, via $Y_{1,2}^*$ (Figs. \ref{ffig3} (a)) along the grey and black trajectories (Figs. \ref{ffig3} (b) and (d)) after scrolling out around the unstable equilibria $X_{3,4}^*$, or tend to $X_{1,2}^*$ along the stable $1D$ manifold of $X_{1,2}^*$ (dotted red and blue trajectories in Figs. \ref{ffig3} (b) and (d)), as $t\rightarrow\infty$.

Next, consider equilibria $X_{3,4}$ which, due to the symmetry (\ref{sim}), behave similarly. Fig. \ref{ffig3} (c), for the sake of clarity, presents only 25 trajectories starting from the vicinity of $X_4^*$. It reveals that the trajectories either tend to infinity by scrolling out around the unstable $1D$ manifold of $X_4^*$ (dark-brown trajectory) or tend to the equilibrium $X_1^*$ along the stable $1D$ manifold of $X_1^*$ (blue trajectory). Similarly, the simplified case of only two representative trajectories starting from $V_{X_3^*}$ (Fig. \ref{ffig3} (e)) indicates that all trajectories starting in the $\delta$-neighborhood of $X_{3}^*$ tend either to infinity (black trajectory) or to $X_2^*$ (dotted red trajectory).

Note that too large vicinities might lead to intersection of the neighborhoods of the equilibria with the basin of attraction of the considered chaotic attractor.

Summarizing, all trajectories starting from the neighborhoods of unstable equilibria $X_{0,3,4}$, either tend to infinity, or are attracted to the stable equilibria $X_{1,2}^*$ but are not attracted by the chaotic attractor. This numerical analysis leads to a conclusion that the chaotic attractors obtained in system (\ref{rf}) are very likely to be hidden.

\begin{figure}
\begin{center}
  \includegraphics[clip,width=0.8\textwidth] {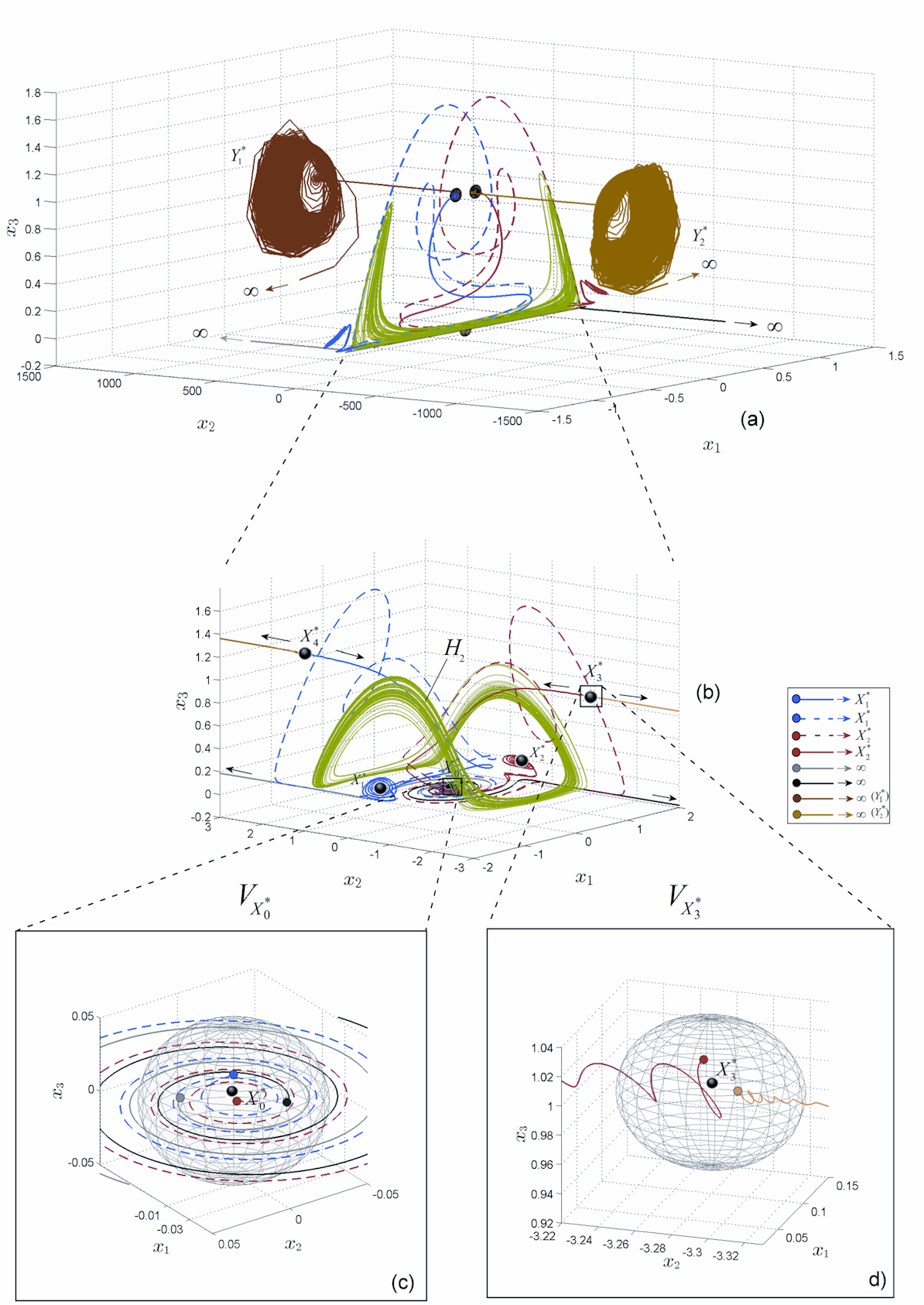}
\caption{(a) Hidden chaotic attractor (green) for $b=0.2715$ and the ``virtual'' saddles $Y_{1,2}^*$. (b) Detail of the hidden chaotic attractors, trajectories diverging to infinity (via $Y_{1,2}^*$) and trajectories attracted by equilibria $X_{1,2}^*$. (c) Trajectories starting from the $\delta$-vicinity of $X_0^*$, either diverge to infinity via $Y_{1,2}^*$ (black and grey trajectories), or tend to $X_{1,2}^*$ (dotted red and blue trajectories) (4 representative trajectories). (d) Trajectories starting form the $\delta$-vicinity of $X_3^*$ either tend to $X_2^*$ (red), or diverges to infinity via $Y_2^*$ (brawn) (two representative trajectories).}
\label{ffig4}
\end{center}
\end{figure}

Another hidden chaotic attractor $H_2$ corresponds to $a=0.1$ and $b=0.2715$ (see Fig. \ref{ffig4}). The difference from the case of $b=0.2876$ consists in the fact that the divergent trajectories starting from $V_{X_0^*}$ are no longer attracted by the ``virtual'' saddles (Fig. \ref{ffig4} (a), (b)).

\section{Conclusion}
In this paper, it is shown that, compared to many other classical nonlinear systems, the RF system presents unusual dynamics (e.g. ``virtual'' saddles), which appear for a large interval of the parameter $b$. Also, hidden chaotic attractors have been identified. Contrarily to the common rule that numerical results should be independent of the numerical methods and the integration step size if the methods and steps are used properly, the results obtained in the case of the RF system depend on the utilized numerical integration method and the step size; otherwise, some intrinsic interesting dynamical behavior will not be able to find.

\end{document}